\documentclass[twocolumn]{aastex62}
\usepackage[T1]{fontenc}
\usepackage[figure,figure*]{hypcap}
\shorttitle{CEMP-no Group Morphology in the Milky Way}
\shortauthors{Yoon et al.}
\begin{document}

\title{Origin of the CEMP-no Group Morphology in the Milky Way}
%\title{Origin of the CEMP-no Group Morphology in the Milky Way in the Context of Hierachical Assembly of the Galactic Halo}

\author[0000-0002-4168-239X]{Jinmi Yoon}
\affiliation{Department of Physics and JINA-CEE, University of Notre Dame, Notre Dame, IN 46556}
\correspondingauthor{Jinmi Yoon}
\email{jinmi.yoon@nd.edu}

\author[0000-0003-4573-6233]{Timothy C. Beers}
\affiliation{Department of Physics and JINA-CEE, University of Notre Dame, Notre Dame, IN 46556}

\author{Di Tian}
\affiliation{Department of Physics, Xi'an Jiaotong University, Shaanxi, 710049, People's Republic of China}

%\altaffiltext{REU student at University of Notre Dame}
\author[0000-0002-9594-6143]{Devin Whitten}
\affiliation{Department of Physics and JINA-CEE, University of Notre Dame, Notre Dame, IN 46556}

\begin{abstract}

The elemental-abundance signatures of the very first stars are imprinted on the atmospheres of CEMP-no stars, as various evidence suggests they are bona-fide second-generation stars. It has recently been recognized that the CEMP-no stars can be sub-divided into at least two groups, based on their distinct morphology in the $A$(C)-[Fe/H] space, indicating the likely existence of multiple pathways for their formation. In this work, we compare the halo CEMP-no group morphology with that of stars found in satellite dwarf galaxies of the Milky Way -- a very similar $A$(C)-[Fe/H] pattern is found, providing clear evidence that halo CEMP-no stars were indeed accreted from their host mini-halos, similar in nature to those that formed in presently observed ultra faint dwarfs (UFDs) and dwarf spheroidal (dSph) galaxies.  We also infer that the previously noted ``anomalous'' CEMP-no halo stars (with high $A$(C) and low [Ba/Fe] ratios) that otherwise would be associated with Group I may have the same origin as the  Group III CEMP-no halo stars, by analogy with the location of several Group III CEMP-no stars in the UFDs and dSphs and their distinct separation from that of the CEMP-$s$ stars in the $A$(Ba)-$A$(C) space. Interestingly, CEMP-no stars associated with UFDs include both Group II and Group III stars, while the more massive dSphs appear to have only Group II stars. We conclude that understanding the origin of the CEMP-no halo stars requires knowledge of the masses of their parent mini-halos, which is related to the amount of carbon dilution prior to star formation, in addition to the nature of their nucleosynthetic origin.

\end{abstract}

\section{Introduction}\label{intro}

One of the primary goals of Galactic Archaeology is to understand the nature of the first stars and their contribution to the chemical enrichment of the early Universe and early galaxy formation. The first stars are thought to be massive and short-lived, which enriched their surrounding pristine gas soon after their birth. Their nucleosynthetic imprints are locked up in the next-generation stars, so we are able to understand the first stars through indirect studies of their immediate descendants, the so-called CEMP-no stars (see the summary of evidence presented in section 4.3 of \citealt{hansen2016a}). CEMP-no stars are a sub-class of carbon-enhanced metal-poor (CEMP) stars\footnote{There are several CEMP sub-classes, depending on enhancement of heavy neutron-capture elements such as Ba and Eu. CEMP-$s$ : [C/Fe] $\geq$ +0.7 and [Ba/Fe] $>$ +1.0 and
[Ba/Eu] $>$ +0.5, CEMP-$r$ : [C/Fe] $\geq$ +0.7 and [Eu/Fe]$>$ +1.0, 
CEMP-$r/s$ or CEMP-$i$ : [C/Fe] $\geq$ +0.7 and 0.0 $<$ [Ba/Eu] $<$+0.5, CEMP-no : [C/Fe] $\geq$ +0.7 and [Ba/Fe] $\leq$ 0.0.  Here we indicate relative abundance ratios as [A/B]= log $(\rm N_A/N_B)_{\star}$ - log $(\rm N_A/N_B)_{\odot}$ where $\rm N_A$ and $\rm N_B$ are numbers of elements A and B respectively.} \citep{beers2005,aoki2007} that exhibit strongly enhanced carbon but low abundances of heavy neutron-capture elements. There is increasing evidence that many of them, in particular, the extremely metal-poor ([Fe/H] $<-$3.0) CEMP-no stars, exhibit enhancements of light elements such as N, O, Na, and Mg relative to Fe as well \citep[e.g.,][]{yong2013,aoki2018}. 

Many studies have attempted to identify the possible progenitors of the CEMP-no stars by matching their elemental-abundance patterns to those of theoretical models employing various nucleosynthetic sites/processes such as mixing and fallback ``faint'' supernovae (faint SNe) of the first massive stars \citep[e.g.,][]{umeda2003,nomoto2013,tominaga2014}, extremely metal-poor rapidly-rotating massive stars  \citep[``spinstars'', e.g.,][]{meynet2006,meynet2010,chiappini2013,maeder2015, frischknecht2012,frischknecht2016,choplin2017}, and/or intermediate neutron-capture processes associated with high-mass early-generation stars \citep{clarkson2018}.  

\begin{figure*}
\includegraphics[scale=0.8]{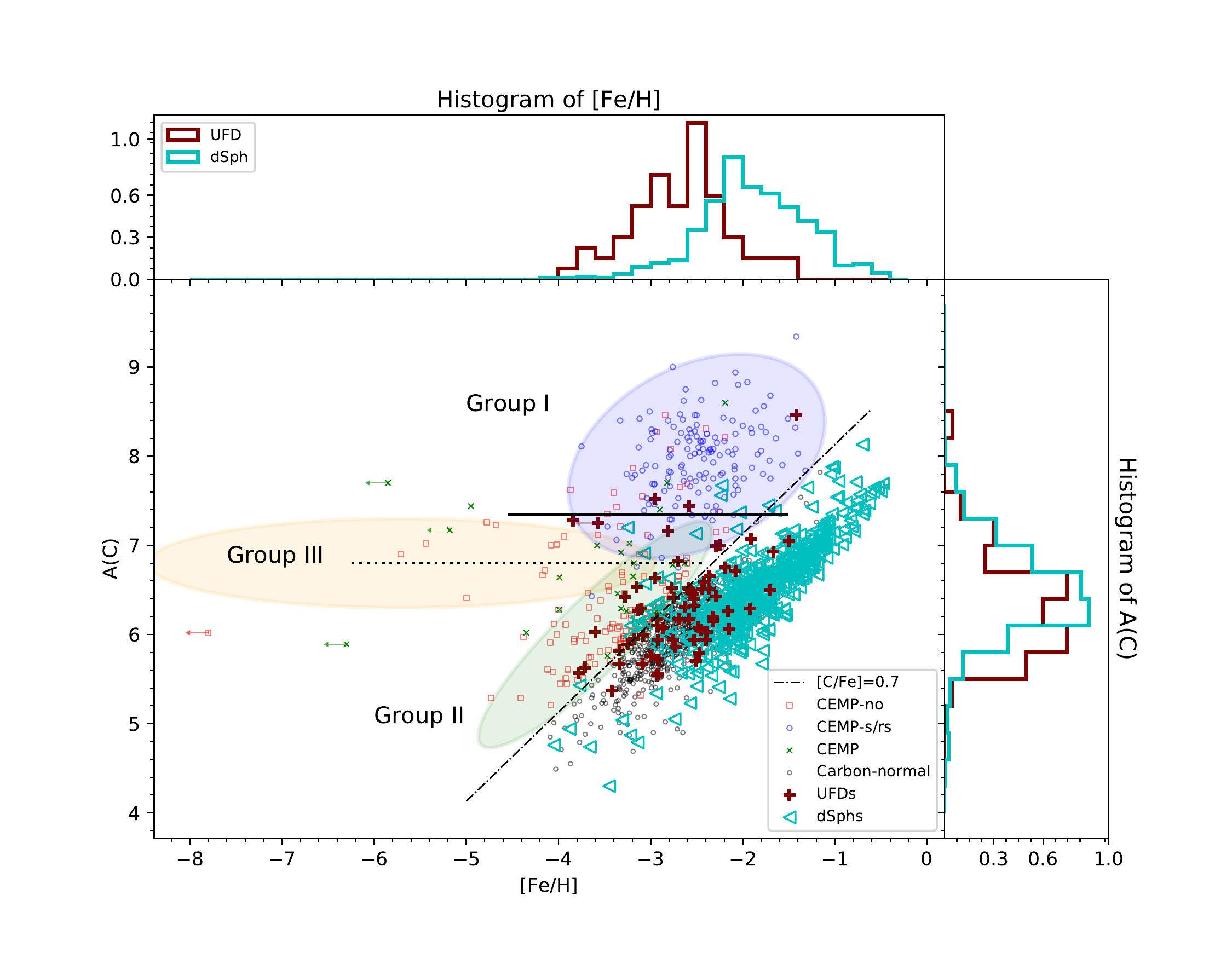}
\caption{The $A$(C)-[Fe/H] diagram of stars from the dSphs (cyan triangles) and UFDs (maroon + signs), along with the halo CEMP stars (CEMP-$s/rs$: blue  circles, CEMP-no: red squares, CEMP stars without Ba detection: green crosses, and C-normal stars: gray circles). The black-dotted and solid-horizontal lines represent average $A$(C) levels, estimated by eye, for the halo Group III stars and the satellite galaxies, respectively. The marginal plots show histograms of $A$(C) and [Fe/H]. The cyan and maroon histograms indicate the dSph and UFD stars, respectively. The orange, green, and blue ellipses  correspond to suggested divisions for the CEMP groups, adopted from \citet{yoon2016}.
\label{ybd}}
\end{figure*}

The latest challenge for understanding the origin of the halo CEMP-no stars emerged from close inspection of the three morphologically distinct CEMP groups identified in the Yoon-Beers diagram of  $A$(C)\footnote{$A$(X)=$\log\,\epsilon$(X)=$\log\,$($N_{\rm X}/N_{\rm H}$)+12, where $N_{\rm X}$ and $N_{\rm H}$ represent number-density fractions of a given element X and hydrogen, respectively.}-[Fe/H], shown in Figure~\ref{ybd}. This distinct division indicates that multiple origins are likely responsible for the apparent CEMP groups \citep{yoon2016}. According to Yoon et al., a subset ($\sim$13\%) of the CEMP-no stars reside in the Group I region, which is predominantly occupied by the CEMP-$s$ stars, whose strong carbon and barium enhancement is thought to arise due to external pollution associated with mass-transfer from a binary companion during its asymptotic giant branch (AGB) phase {\citep[e.g.,\bf][]{suda2004,herwig2005,lucatello2005b,bisterzo2011,abate2013, starkenburg2014,abate2018}}.

In contrast, the majority of the CEMP-no stars are located in the Group II and III regions, with substantially lower $A$(C) and [Fe/H] than the Group I stars. The Group II and Group III stars show clearly different behaviors as well -- the $A$(C) of the Group II stars is correlated with [Fe/H], while that of the Group III stars
appears to have no strong dependence on [Fe/H]. In addition, the Group III stars have $A$(C) $ \sim 1-2$ dex higher, on average, than that of the ultra metal-poor ([Fe/H] $ <-4.0$) Group II stars. As further shown by \citet{yoon2016}, the Group II and III stars also exhibit distinctly different trends in the space of absolute abundances of Na and Mg, $A$(Na, Mg), as a function of $A$(C).

Most studies of the origin of CEMP-no stars have focused on their astrophysical progenitors and the mass ranges associated with these progenitors \citep[e.g.,][]{placco2016b, ishigaki2018}, traced by the observed abundance patterns, the possibility of binary mass-transfer origin \citep[e.g.,][]{suda2004,arentsen2019}, the cooling channels through which low-mass second-generation stars might form \citep{chiaki2017},  formation as the result of inhomogeneous mixing \citep{hartwig2019}, and the nature of the pollution event(s), either external or internal to their natal clouds \citep{chiaki2018,chiaki2019}. However, the distinct CEMP-no groups identified in the absolute abundance space of $A$(C) likely also reflect, at least to some extent, the dilution level of the first-star nucleosynthesis products due to mixing with the pristine gas available in their birth environment. In this work, we focus on investigating whether the different CEMP-no groups in the halo could be related to the Galactic assembly process, in the context of the role of the baryonic mass of their host mini-halos, by comparing the $A$(C)-[Fe/H] morphology of the CEMP-no stars associated with the satellite dwarf galaxies of the Milky Way (MW).

Although our primary focus in this work is on the Galactic accretion history of CEMP-no stars, we also recognize the importance of the most significant elements (C, Fe, and Ba) used to characterize CEMP-no stars, in the context of the interwoven relationships of the star-formation history (SFH), stellar populations, Galactic chemical evolution, and halo-assembly history. Thus, in Section ~\ref{origin}, we provide a brief summary of the origin and cosmic-evolution history of carbon, iron, and barium, in order to to facilitate our later inferences. We describe the literature data compilation in Section~\ref{data}. In Section~\ref{sp},
we show the $A$(C)-[Fe/H] distribution of stars from the satellite dwarf galaxies, and consider the implications of the nature of their stellar populations in the context of SFHs in different galactic environments. In Section~\ref{group1} we provide evidence that the halo Group I CEMP-no stars might have the same nucleosynthetic origin as the Group III CEMP-no stars.  We discuss the accretion history of the halo CEMP-no stars in Section~\ref{accretion}, and the role of host mini-halo masses on the CEMP-no groups in Section~\ref{galaxymass}. We summarize our conclusions in Section~\ref{conclusion}. 

\section{The origin and evolution of carbon, iron, and barium}\label{origin}
Carbon, iron, and barium are the ``first-approximation'' distinctive elements used for the identification of CEMP-no stars \citep{beers2005}. They play important roles in our understanding of early Galactic chemical evolution through their own nucleosynthetic origin and evolution history, and also serve as tracers of stellar populations and SFHs. In particular, each element has two dominant pathways characterized by different stellar mass scales. One pathway transitions into the other as the dominant chemical-enrichment channel -- contributions from massive stars give way to that from intermediate- and low-mass stars over cosmic time \citep[e.g.,][and references therein]{frebel2013b,frebel2015,yoon2018}.

Carbon and iron are synthesized during the core helium and advanced burning stages (O, Ne, and Si), respectively, in the central regions of the first massive stars. Carbon is ejected via faint SNe, normal core-collapse SNe (CCSNe), and/or mass-loss of spinstars and iron through predominantly CCSNe. Roughly a Gyr later, the main production channel of carbon transitions to low- and intermediate-mass thermally pulsating AGB stars, accompanied by activation of the slow neutron-capture process (main $s$-process) \citep[e.g.,][]{busso1999,travaglio2004, herwig2005,karakas2014}. Carbon is ejected via stellar winds or, under certain circumstances, accreted into the atmospheres of their less-evolved binary companions (CEMP-$s$ stars). Similarly, the dominant Fe production pathway changes from massive stars to the significantly larger contribution from Type Ia SNe (SNIa), believed to originate from low- to intermediate-mass stellar binary mass transfer or double-degenerate mergers \citep[e.g.,][and references therein]{tolstoy2009}.

The dual production pathways of carbon and the transition between the above channels are revealed by the bi-modal $A$(C) distribution \citep{spite2013,bonifacio2015,yoon2016,lee2017} of the halo CEMP stars. The transition of the iron-production channel from massive stars to lower mass stars is thought to account for the so-called ``knee'' in the $\mathrm{[\alpha/Fe]}$-[Fe/H] space, at which $\alpha$-element over-abundances with respect to Fe begins declining to the Solar ratio. The position of the knee indicates the onset of the dominant contribution from SNIa -- in the MW at about [Fe/H] $\sim -1.0$ \citep{tolstoy2009}. The location of the knee in dwarf galaxies varies over a wide range ($-3.0< \mathrm{[Fe/H]} <-1.0$)  due to their individual star-formation and enrichment histories. Some of the UFDs do not show a clear knee due to the small number of stars observed \citep[e.g,][]{kirby2010, frebel2012,simon2019}.

According to current understanding, the main astrophysical production site of Ba in the early Universe is likely related to massive stars via a rapid neutron-capture process ($r$-process) \citep{pagel1997,travaglio1999,cescutti2006}{\bf. Photometric and spectroscopic bservations of the electromagnetic counterpart \citep[AT 2017gfo;][]{cowperthwaite2017,drout2017,kilpatrick2017,shappee2017} of the recent LIGO/VIRGO detection of a binary neutron star merger in 2017 \citep[GW170817;][]{abbott2017a,abbott2017b} provided direct evidence of  main $r$-process element production. An additional source, operation of an $s$-process in spinstars, has been recently suggested \citep{frischknecht2012,cescutti2013,maeder2015b,frischknecht2016, choplin2017,choplin2018}, which could also possibly produce Ba. Following these,} the dominant site for Ba (along with carbon) production transitions to low- to intermediate-mass thermally pulsating AGB stars, where the main $s$-process operates. This shift accounts for the transition from the relative dominance of CEMP-no stars to CEMP-$s$ stars in the individual distributions of [Fe/H] and $A$(C) shown in Figure 2 of \citet{yoon2016}, and indirectly through the bi-modal $A$(C) distribution shown in the Yoon-Beers diagram.  

Hence, when considering the origin and evolution history of these critical elements, each element is used to probe star formation on similar timescales to the other elements, and as tracers of the SFH and contributions from different stellar populations. Furthermore, consideration of the spatial distributions of these elemental abundances, in particular, C and Fe, in the MW halo could provide constraints on the assembly history of the Galactic halo \citep{lee2017, yoon2018}.

\begin{figure*}
\includegraphics[scale=0.6]{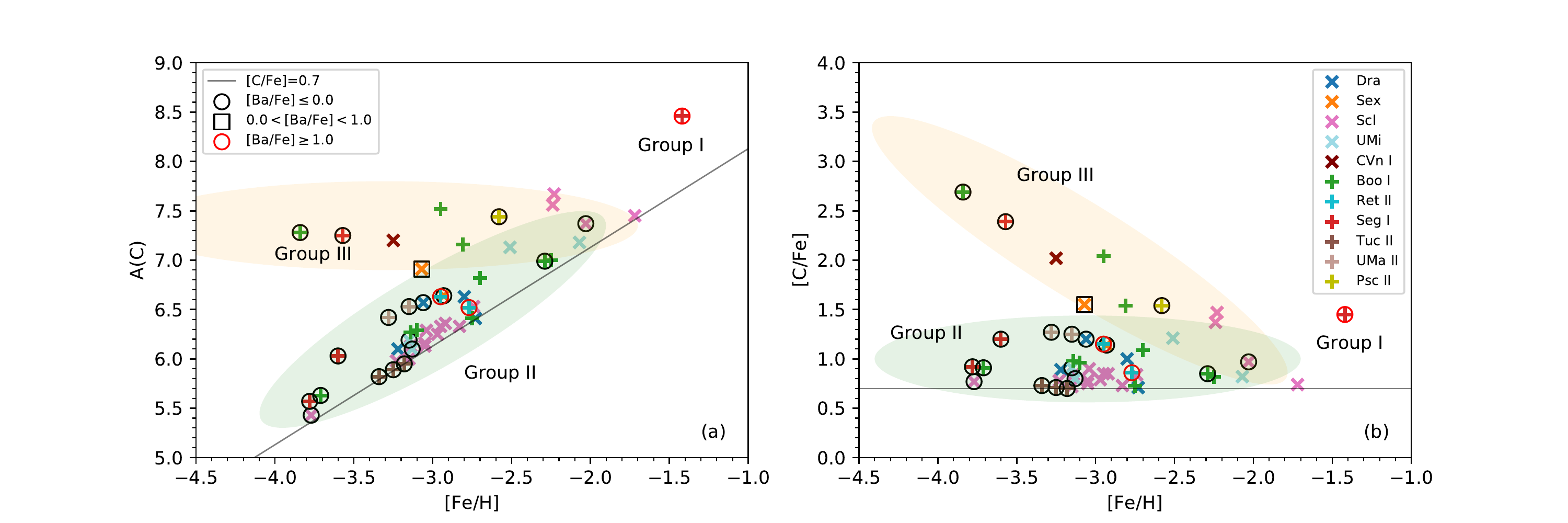}\\
\includegraphics[scale=0.6]{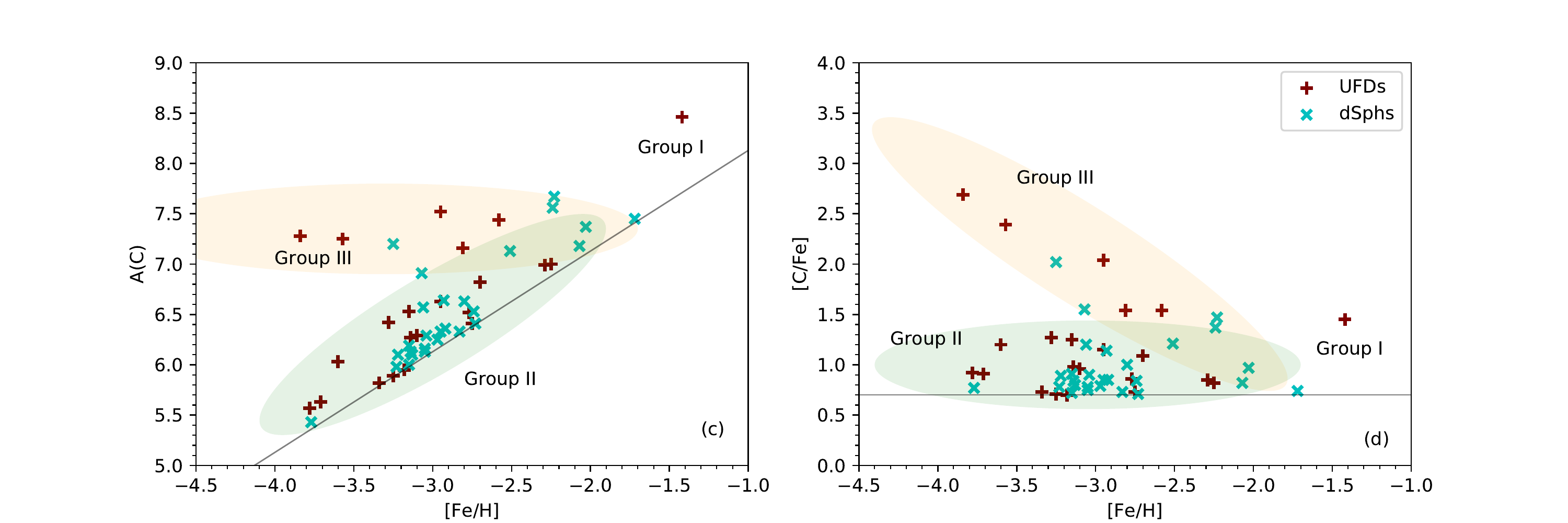}
\caption{ The CEMP distribution of the UFDs (+ signs) and dSphs (x signs) in  the $A$(C)-[Fe/H] and [C/Fe]-[Fe/H] spaces. The gray lines in the panels represent the criterion for CEMP stars ([C/Fe] $=+0.7$). The orange and green ellipses represent the suggested CEMP groups. The symbols in the legend of panel (a) indicate the level of barium enhancement, while the host galaxies of the stars are shown in the legend of panel (b). Panels (c) and (d) show the contrast between the stars from the UFDs (maroon + signs) and dSphs (cyan x signs). The original references of abundances are \citet{cohen2010,frebel2010,tafelmeyer2010,honda2011, lai2011,shetrone2013,frebel2014,kirby2015,skuladottir2015,ji2016d,lardo2016,chiti2018a,chiti2018b, spite2018}.
\label{ybd2}}
\end{figure*}

\section{Literature Data}\label{data}
In order to carry out a comparison of the morphology of CEMP groups in the  $A$(C)-[Fe/H] space for halo stars and stars in satellite galaxies, we first assemble a literature sample from a number of sources. We then applied the carbon correction calculator\footnote{\url{http://vplacco.pythonanywhere.com/}} of \citet{placco2014c} to the measured $A$(C) for all stars in the sample, in order to obtain an estimate of the natal absolute carbon abundance, prior to dilution during late giant-branch evolution. The definition of CEMP stars is taken as [C/Fe] $\geq +0.7$, after applying the carbon-evolution correction. Relative abundances were derived by using the Solar abundances of \citet{asplund2009}. 

A variety of spectral resolving powers have been used to identify C-normal and CEMP stars in the literature, ranging from medium- ($R \sim 2000$), to moderate- ($5000 \leq R \leq 10000$), to high-resolution spectroscopy ($R \geq 20,000$).   For the halo stars, the C-normal metal-poor stars are collected from \citet{placco2014c}. and the CEMP stars from \citet{yoon2016}, both based on predominantly high-resolution spectroscopy (although several stars had only moderate-resolution spectroscopy available). The CEMP stars from recently published moderate- \citep{aguado2018a,bonifacio2018} and high-resolution spectrosocopic studies \citep{aoki2018,frebel2019} are included as well. The halo CEMP-$r$ stars\footnote{CEMP-$r$ stars are the rarest stellar population among the CEMP stars, whose astrophysical origin include rapid neutron-capture process sites, and are also thought to be accreted into the halo \citep{ji2016b}. Their low $A$(C) is expected, because the CEMP-$r$ stars are not apparently associated with binary systems \citep{hansen2015b}, though their [Ba/Fe] is relatively higher than those of CEMP-no stars.} are not included.

For the satellite galaxies, the majority of the C-normal and CEMP stars are retrieved from the SAGA database\footnote{\url{http://sagadatabase.jp}} \citep{suda2008}, based on moderate- to high-resolution spectroscopy.  The 11 recently confirmed CEMP stars from Sculptor (Scl) \citep[with medium-resolution spectroscopy,][]{chiti2018a}, along with three CEMP-no stars from Tucana II (Tuc II) \citep[with high-resolution spectroscopy,][]{chiti2018b}, and another from Pisces II (Psc II) \citep[with moderate-resolution spectroscopy,][]{spite2018} are included, as well as a recent medium-resolution spectroscopic result for one carbon giant in Canes Venatici I (CVn I) \citep{yoon2019b}. The final sample only includes stars with carbon detections for classical dwarf spheroidal galaxies (dSphs) and ultra-faint dwarf galaxies (UFDs). Hence, the fraction of stars (the number of stars with carbon measurement/total observed stars) in each dwarf galaxy analyzed in this work differs from galaxy to galaxy: 
For the dSphs -- CVn I (1/201), Draco (Dra: 158/346), Fornax (46/1400), Scl (293/609), Ursa Minor (UMi: 51/226), Sextan (Sex: 3/301), and for the UFDs -- Bootis I (Boo I: 31/63), Bootis II (4/6), Coma Berenices (3/9), Hercules (1/30), Reticulum II (Ret II: 8/9),  Segue I (Seg I: 8/11), Triangulum II (1/15), Tuc II (7/7), Ursa Major II (UMa II: 3/17), and Psc (1/7). Since many UFDs have only a very small number of stars with reported spectroscopic results, we only consider the global SFHs of dwarf galaxies in terms of the galaxy type (dSphs or UFDs) collectively, rather than for individual galaxies.

We note that the halo stars considered in this work are predominantly ($\sim$70\%) subgiant or giant stars (log~$g < 3.5$). Therefore, any bias due to different luminosity classes on our inferences in the following sections is likely to be insignificant.

\section{CEMP Group Morphology for Stars in MW Satellite Dwarf Galaxies}\label{sp}

\subsection{The Stellar Populations in MW Satellite Dwarf Galaxies} 

Figure~\ref{ybd} shows the updated version of the Yoon-Beers  $A$(C)-[Fe/H] diagram \citep{yoon2016}. The new additions are the metal-poor stars from the dwarf galaxies and the halo C-normal stars, along with halo CEMP stars without reported barium abundances, which were not included in the original diagram.

From inspection of Figure~\ref{ybd}, it is clear that the dwarf galaxy stars are distributed similarly to the halo stars. The metallicity distribution function (MDF) of the canonical dSph stars is significantly shifted toward higher metallicities compared with that of the UFDs, as seen in the top marginal plot. This shift in [Fe/H] can be explained by the different level of gas available for star formation in each system: Massive dSphs (total mass $\gtrsim 10^9{\rm M}_\odot$) have much higher gas content than the UFDs (total mass $\lesssim 10^6{\rm M}_\odot$), resulting in more extended episodes of star formation, and more metal (iron)-rich environments, as normal CCSNe are the predominant source of iron. The UFDs are likely to have truncated star formation, due to the limited supply of gas and/or stellar feedback from reionization, thus faint SNe, which produce very little iron, were likely the main sources for their metal enrichment, resulting in much more metal (iron)-poor environments \citep[][and references therein]{mateo1998,tolstoy2009,walker2009,frebel2012,mcconnachie2012,frebel2015b, salvadori2015,jeon2017,simon2019}. In particular, the contrast in the MDFs and SFHs between the UFDs and dSphs is consistent with theoretical predictions  (\citealt{salvadori2015}, their Figures 3 and 4). The most luminous (corresponding to more-massive dSphs) dwarf galaxies have extended SFHs ($0<z<13$, peaking at redshift $z \lesssim5$) with high star-formation rates at later time ($z\sim 0$), and, in turn, relatively narrow MDFs ($-2.0 < \mathrm{[Fe/H]}<0.0$), peaking at $-2.0< \mathrm{[Fe/H]} < -1.0$. In contrast, the least luminous (UFD-like) galaxies have dominant high-redshift SFHs ($7<z<13$), peaking at $z\sim 10$, that are truncated before reionization ($z\sim 6$).  As a result, they exhibit extended MDFs ($-5.0 < \mathrm{[Fe/H]}<-1.0$) with peaks shifted toward much lower metallicity.

\begin{figure*}
\centering
\includegraphics[scale=0.9]{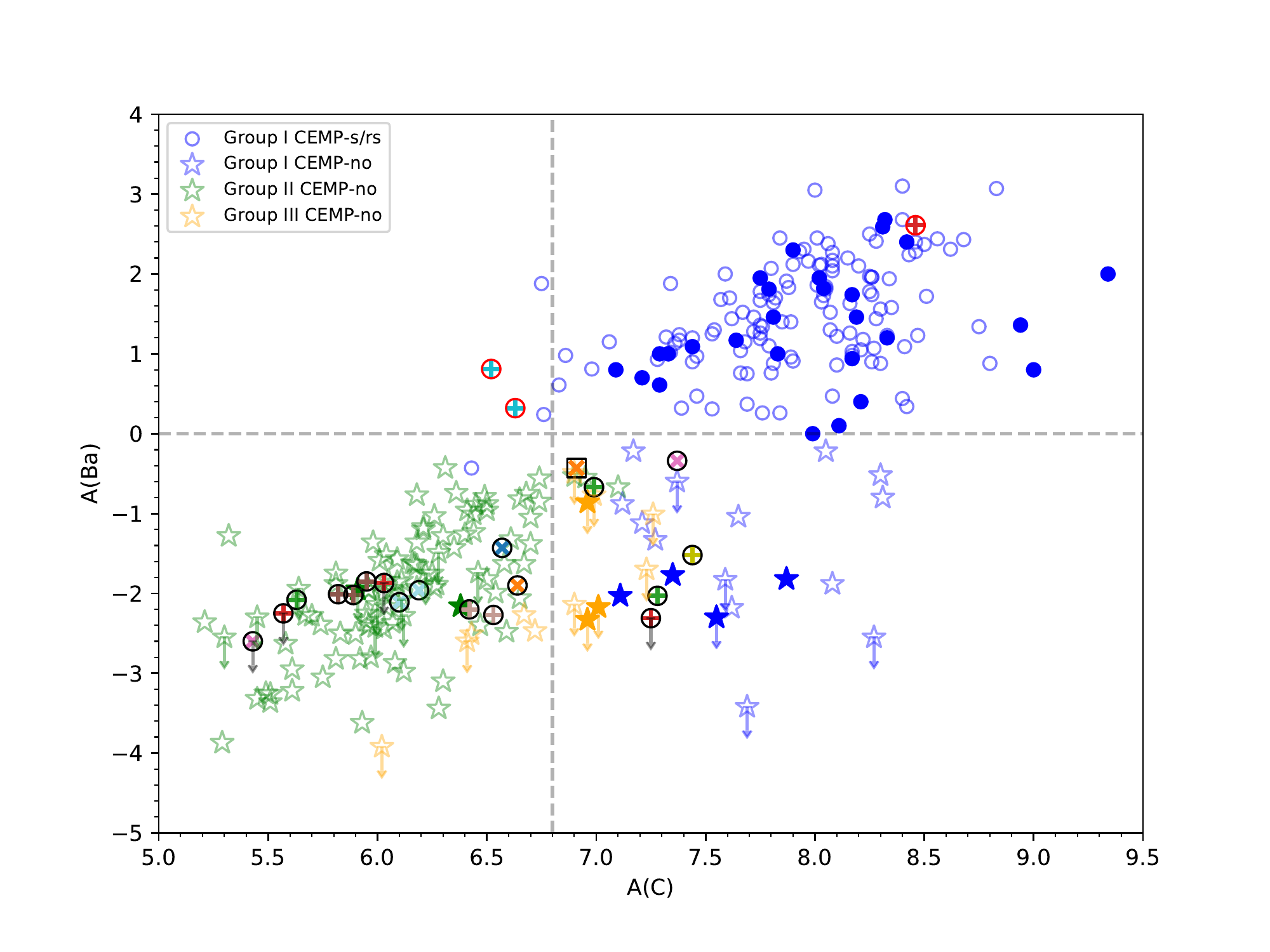}
\caption{$A$(Ba) vs. $A$(C) for CEMP stars from MW halo and dwarf galaxies. Halo CEMP stars are labeled in the legend at the upper left; the CEMP stars in dwarf galaxies are plotted using the same symbols shown in Figure~\ref{ybd2}. Filled symbols represent binary stars reported  from \citet{yoon2016} and \citet{arentsen2019}. Upper limits on Ba are plotted as downward arrows. We note that two stars from Ret II are CEMP-$r$ stars, seen in the upper-left quadrant in the figure.  One Group II CEMP-no binary star is not visible because it is crowded with three stars from Tuc II. The dashed vertical and horizontal lines are plotted, for convenience, to guide our discussion in the text. \label{aba_ac}}
\end{figure*}

It is interesting to note that the $A$(C) histograms of the UFDs and the dSphs are quite similar, as shown in the right marginal plot in Figure~\ref{ybd}, suggesting that the available carbon-production channels and the evolution of carbon are universal, and do not rely significantly on galactic environment. We note that contributions of carbon from AGB stars is not clearly seen in Figure~\ref{ybd}, unlike that for the halo stars.  In fact, there is only one CEMP-$s$ star, likely polluted by an AGB companion \citep{frebel2014}, reported among the dwarf galaxies. However, the C-normal stars with $A$(C) $> 7.1$ with measured [Ba/Fe] (only several stars exist) from the dwarf galaxies, are found in the region of the halo CEMP-$s$ stars in the $A$(Ba)-$A$(C) space (not shown in Figure~\ref{aba_ac}). This indicates that the $s$-process enabled enrichment of the galactic environments of dSphs up to Solar C and Ba values. However, their metallicity is [Fe/H] $ >-1.5$, thus we suspect that the lack of CEMP-$s$ stars among the dwarf galaxies arises from the aforementioned SFHs (resulting in dilution of CEMP signatures, as discussed below) rather than from an observational bias.

From the above, it is evident that there exists a correlation between the dominant stellar populations and CEMP fractions and the masses of their host galaxies  \citep[e.g.,][]{norris2010b,frebel2014,salvadori2015}. For the dSphs, only a small fraction ($\sim 3$\%) of the metal-poor stars are CEMP stars, since the higher iron production (due to prolonged SFHs and dominant contributions of CCSNe and SNIa) relative to carbon in the system dilutes the CEMP signature. In contrast, a substantial fraction ($\sim$ 28\%) of the metal-poor stars from the UFDs are CEMP stars (due to the relatively lower iron production arising from the truncated SFHs and dominant faint SNe contribution at early times) compared with carbon production in their natal system, enabling preservation of the CEMP signature, similar to the arguments put forward previously by \citet{salvadori2015} and \citet{skuladottir2015}.

%\vfill\eject

\subsection{CEMP Sub-classes in the Satellite Galaxies}

The sub-classes of CEMP stars associated with specific astrophysical production sites (such as faint SNe, and sites of the $s$-process and $r$-process) were first suggested as a standard taxonomy in \citet{beers2005}, with the caution that these divisions are somewhat arbitrary due to the existence of ``... continuous, rather than discrete, distribution of the relevant parameters''. Hence, the nomenclature should be considered as a ``first approximation'', subject to future revision. Since then, various studies \citep[e.g.,][]{aoki2007,norris2010, bonifacio2015, maeder2015,hampel2016, yoon2016,chansen2019} have suggested various revisions for the sub-classes, as additional data have been obtained, and better understanding of the nature of these stars has emerged.

\citet{yoon2016} suggested an approach to distinguish CEMP-no stars from CEMP-$s$ stars based solely on the absolute level of carbon abundance, $A$(C), as the bi-modal (low and high peaks) of the $A$(C) distribution corresponds well to the expected {\it intrinsic} (natal) and {\it extrinsic} (mass-transfer) nature of the CEMP-no and CEMP-$s$ stars, respectively. They further demonstrated that this approach is as effective ($\sim$90\%) as the conventional method, which makes use of the [Ba/Fe] ratio (and requires high-resolution spectroscopy to obtain a measurement of Ba). This effectiveness can be validated from the overall abundance trends of C, Fe, and Ba, which are well-correlated with one another, as seen in Figure 6 and 7 in \citet{yoon2016} and Figure~\ref{aba_ac} in this work. This method is especially useful when only medium-resolution spectroscopy is available (or if there is no Ba measurement). As seen in Figure~\ref{ybd},  there are 27 CEMP stars found in the dSphs (Dra, Sex, Scl, and UMi) and 22 CEMP stars from the UFDs (Boo I, Psc II, Ret II, Seg I, Tuc II, and UMa II). We note that two CEMP stars from Ret II are CEMP-$r$ stars. Twenty-six out of the 49 CEMP stars do not have reported Ba measurements. From application of the $A$(C) method, we find that the majority of the CEMP stars in dSph and UFD galaxies are CEMP-no stars.

Among the satellite-galaxy stars with $A$(C) $> 7.1$, eight CEMP stars have unknown Ba status. Six of them are from the dSphs CVn I, Scl, and UMi I. The other two CEMP stars are from the UFD Boo I.  According to the $A$(C) separation, as previously employed, these stars could all be CEMP-$s$ stars, but we suspect that the majority of these unclassified CEMP stars might in fact be CEMP-no stars, for a number of reasons. First, there exist several halo CEMP-no stars with similar high values of $A$(C). Secondly, few CEMP-$s$ stars in the satellite systems are likely to be found, due to the aforementioned dilution of the CEMP signature, especially at higher metallicity ([Fe/H] $> -$2.0). As mentioned above, only one (relatively metal-rich, [Fe/H]$\sim-1.4$)  CEMP-$s$ star, with high $A$(C) $\sim 8.5$ \citep[in Seg I;][]{frebel2014}, has been reported in the satellite galaxies to date. Thirdly, the UFD stars are predominantly found at extremely low metallicity, where most halo CEMP-no stars are also found. Fourthly, the metallicity range of these CEMP stars, in particular in dSphs, is where many CEMP-no stars are predicted to be found, according to \citet{salvadori2015}. Most importantly, all CEMP stars with known [Ba/Fe] (except two CEMP-$r$ stars from Ret II and one CEMP-$s$ from Seg I, which have high Ba abundances) from the dwarf galaxies are found among the halo CEMP-no stars, and clearly separated from the halo CEMP-$s$ stars in the upper-right quadrant of the $A$(Ba)-$A$(C) space shown in Figure~\ref{aba_ac}. Thus, the CEMP stars with unknown [Ba/Fe] that reside in the Group III (and II) regions are likely CEMP-no stars.

\section{The nucleosynthetic origin of CEMP-no stars}\label{group1}

\citet{yoon2016} used the observed distribution of $A$(C) vs. [Fe/H] for MW halo stars to argue for the existence of at least three distinct morphological groups of CEMP stars -- Group I (predominantly CEMP-$s$) stars, and Groups II and III (CEMP-no) stars.  The latter two groups exhibit clearly different behaviors with respect to their $A$(C) vs. [Fe/H], as well as in their $A$(Na,Mg) vs. $A$(C) spaces, from which these authors inferred the existence of multiple progenitors and enrichment histories for CEMP-no stars in the MW halo.  The known Group II CEMP-no stars greatly outnumber the known Group III stars, probably in part due to the relative rarity of UMP ([Fe/H] $<-4$) stars.

\citet{yoon2016} also pointed out the existence of a number of ``anomalous'' CEMP-no stars, with high $A$(C) but low [Ba/Fe] ratios, which occupied the Group I region in Figure~\ref{ybd}, but offered no explanation for this behavior. In Figure~\ref{aba_ac}, it is interesting to see that many Group I and III CEMP-no stars, whose $A$(C) are much higher than Group II CEMP-no stars, coexist in the same region (lower-right quadrant), clearly separate in their $A$(Ba) distribution from both the Group II CEMP-no stars (in the lower-left quadrant) and, in particular, the CEMP-$s$ stars (in the upper-right quadrant). The stars in the lower-right quadrant share the same trend of high $A$(C), but an apparently random scatter of low $A$(Ba); perhaps they have shared a common nucleosynthetic origin. 

The recent radial velocity monitoring study of \citet{arentsen2019} reported that many ($47^{+15}_{-14}$\%) of the higher $A$(C) ($ > 6.6$) CEMP-no stars are binaries, although they obtained the similar result as \citet{hansen2016a} when only considering the lower $A$(C) CEMP-no stars ($18^{+14}_{-9}$\% vs. $17 \pm 9$\%). They also noted the claim that the close-binary (periods $< 10^4$ days) fraction of low-mass ($0.6-1.5$~M$_{\odot}$) stars increases at decreasing metallicity \citep{moe2018}, and concluded that a binary mass-transfer scenario may apply to at least some of the high $A$(C) CEMP-no stars. However, this explanation is incomplete, as it does not simultaneously account for the apparently low barium abundances of these stars (all have an upper limit\footnote{Upper limits on abundances quoted in the literature indicate non-detections or $3\sigma$ limits \citep{roederer2013}.  We note that many of the halo Group III stars have $T_{\mathrm{eff}} > 6000$\,K, which indicates a higher threshold to detect Ba abundance, as investigated by \citet{roederer2013}. This, in turn, may result in a higher estimated upper limit on [Ba/Fe].  This correlation between their high $T_{\mathrm{eff}}$ and a high upper limit [Ba/Fe], which may result in a systematic effect, as seen in the left panel  of Figure 7 in \citet{arentsen2019}. } $A$(Ba) $<$ 0, distinct from the CEMP-$s$ stars). 

In addition, the C and Ba of the Group I and III CEMP-no stars appear uncorrelated (or anti-correlated, to some extent, when considering only stars with detected $A$(Ba)), as can be appreciated from inspection of the lower-right quadrant of  Figure~\ref{aba_ac}. This suggests, for these stars at least, that the production source of carbon and barium is not likely to be the same, or remains an unrecognized process.  While the results from \citet{moe2018} were limited to stars with [Fe/H] $ > -2.0$, currently no well-understood low-metallicity ([Fe/H] $ < -2.0$) AGB mass-transfer models exist that can account for the disparate $A$(C)-$A$(Ba) patterns of Group I and III CEMP-no stars. 

The lack of a trend between $A$(Ba) and $A$(C) seems is similar to that found for $A$(Na) and $A$(Mg) with respect to $A$(C) for the Group III CEMP-no stars \citep{yoon2016}. 

Recent models of spinstars \citep{frischknecht2012,cescutti2013,maeder2015,frischknecht2016, choplin2018} have suggested that a non-standard $s$-process may operate in rapidly rotating massive stars, which might produce barium. Alternately, plausible massive binary (or multiple system) progenitors, which produce abundant carbon, may end up being neutron star mergers at the end of their lifetimes, which could produce Ba via an $r$-process, prior to the formation of the Group I and/or III CEMP-no stars. If the binary fraction indeed increases strongly with decreasing metallicity, such binary progenitor models might be not only plausible but also relatively common and account for the ubiquitously detected Ba among the majority of the most metal-poor stars \citep{roederer2013}. We note that magneto-rotationally
driven jet supernovae {\bf \citep{winteler2012,nishimura2015, nishimura2017} or  the accretion disk around collapsars \citep{pruet2004,surman2004,siegel2018} might be additional sources of Ba production}.

\begin{deluxetable*}{lccc}
\tabletypesize{\scriptsize}
\tablecaption{Characteristics of Group II and Group III CEMP-no Stars  \label{tbl}}
\tablehead{
\colhead{Characteristic}  &\colhead{Group II} & \colhead{Group III} &\colhead{References}}
\startdata
\multicolumn{4}{c}{Chemical signatures}\\
\hline
$A$(C)-[Fe/H] correlation        & yes                     & no \& higher $A$(C)    & (1), (2)\\
$A$(C)-$A$(Na, Mg) correlation   & yes                     & no \& lower $A$(Na,Mg) & (1)\\
$A$(C)-$A$(Ba) correlation       & yes                     & no \& lower $A$(Ba) & (2) \\
\hline
\multicolumn{4}{c}{Galactic environment}\\
\hline
Galaxy type                      & dSphs (\& UFDs)           & UFDs                   & (2) \\
Galaxy total mass& $\sim 10^{9}{\rm M}_{\odot}$ &$\sim 10^{6}{\rm M}_{\odot}$  &(3), (4), (5), (6)     \\
Star-formation history           & prolonged (chemically evolved)               & truncated (stoschastic, inhomogeneous)             & (6), (7) \\
\hline
\multicolumn{4}{c}{Star-forming environment}\\
\hline
Progenitor SN                    & normal CCSNe            & faint SNe              &(8)\\
Gas cooling agents               & silicate grains         & carbon grains          &(8)\\
Dominant Pop. contribution                & Pop. II                 & Pop. III               &(9), (10), (11) \\ 
Number of progenitors            & multiple (multi-enrichment)  & single (mono-enrichment) & (12)\\
Natal-gas enrichment             & internal pollution (self-enrichment)      & external \& internal pollution &(13), (14){\bf, (15)} \\
\enddata
\tablecomments{Some of the characteristics are drawn from inferences based on the results from the listed references; only the dominant source is emphasized in the table. References: (1) \citet{yoon2016}, (2) this work, (3) \citet{mateo1998}, (4) \citet{walker2009}, (5) \citet{mcconnachie2012}, (6) \citet{simon2019}, (7) \citet{tolstoy2009}, (8) \citet{chiaki2017}, (9) \citet{debennassuti2017}, (10) \citet{sarmento2017}, (11) \citet{sarmento2019}, (12) \citet{hartwig2018}, (13) \citet{smith2015}, (14) \citet{chiaki2018}{\bf, (15) \citet{chiaki2019}.}}
\end{deluxetable*}

\section{The Accretion Origin of CEMP-no Stars in the MW Halo}\label{accretion}

With the evolutionary history of carbon and iron in mind, it is clear to see that the group morphology of CEMP stars in the satellite galaxies is very similar to that of the halo CEMP stars, as shown in Figure~\ref{ybd2}. The Group II and Group III regions appear to be clearly divided at [Fe/H] $< -3.0$, as depicted in the green and orange ellipses, respectively. The Group I region has only one known CEMP star, with high $A$(C) (a CEMP-$s$ star from Seg I). The similarity of this morphology is strong evidence that the majority of the halo CEMP-no stars are likely accreted stars that originally formed in environments like the UFD (and some dSph) galaxies. 

%\citet{yoon2016} pointed out the existence of a number of ``anomalous'' CEMP-no stars, with high $A$(C) but low [Ba/Fe] ratios, which occupied the Group I region in Figure~\ref{ybd}, but offered no explanation for this behavior. 
The halo Group I CEMP-no stars of unusually higher $A$(C) seem to be associated with the halo Group III stars, based on inspection of Figure~\ref{ybd}: The distribution of the dwarf galaxy Group III stars are scattered between the halo Group I and the Group III regions, as if they were a part of one larger population encompassing both regions. 
It is interesting to note that there are CEMP-no stars in the satellite galaxies that occupy this same region, suggesting that the original Group III region should perhaps be extended to higher $A$(C) in this part of the $A$(C) vs. [Fe/H] space. This connection between the Group I and III CEMP-no stars is also clearly supported by their co-existence in Figure~\ref{aba_ac}, as discussed in the previous section. If this hypothesis is correct, many of the halo Group I CEMP-no stars, scattered around $A$(C) = 7.4 in Figure \ref{ybd}, may share a common accretion origin with that of the halo Group III CEMP-no stars, as well as a presumably common nucleosynthetic origin. 

As clearly seen in Figure~\ref{ybd2}, the Group II stars are found in both the dSphs and UFDs, while the Group III stars (in particular, at [Fe/H] $\lesssim-$2.5) appear to be found only in the UFDs (with an exception of one star from the CVn I galaxy, presently classified as a dSph). We note that it is not surprising that a Group III star is found in the more massive dSph system CVn I, since CVn I likely has a complex SFH \citep[see, e.g.,][]{ibata2006, martin2008}.  Whatever the astrophysical progenitors of the Group II and III stars are, the UFDs would likely retain the signatures of both groups of stars. In contrast, since the dSphs are clearly chemically evolved systems with prolonged SFHs, their metal-rich environments are likely to have diluted the signature of Group III CEMP-no stars, even if they existed. If this inference is valid, the majority of the Group II CEMP-no stars in the halo were likely accreted from more massive, metal-richer, dSph-like satellite galaxies, whereas the Group III CEMP-no stars were likely accreted primarily from low-mass UFD-like systems.

\section{The Impact of Host Galaxy Environment on CEMP-no Group Morphology}\label{galaxymass}

Table~\ref{tbl} summarizes a comparison of the characteristics of Group II and Group III CEMP-no stars, based on inferences from the latest studies. These two groups of CEMP-no stars have very different properties, with respect to chemical-abundance signatures, their plausible host environments, and, in turn, their overall enrichment and SFHs. 

The differences in the abundances of C, Na, Mg, and Fe between the two CEMP-no groups support the idea that Group II stars likely formed out of natal gas cooled via predominantly silicate grains, such as enstatite (MgSiO$_3$) and forsterite (Mg$_2$SiO$_4$), whose key elements Mg and Si are the main nucleosynthetic products of CCSNe. In contrast, carbon grains, whose key element C is associated with production by faint SNe and/or spinstars, are likely responsible for formation of the Group III CEMP-no stars \citep{chiaki2017}. 

The observed characteristics of each CEMP group appear to correspond well with the nature of the mini-halo environments where they might have formed. UFD-like systems are the least massive dark-matter dominated mini-halos, thus their star formation would be truncated after one or at most a few generations of star formation (depending on their baryonic gas available), due to low star-formation efficiency \citep[e.g.,][]{vincenzo2014} and reionization/SNe feedback \citep[e.g.,][]{salvadori2009,jeon2017}. The most primitive stars in these systems are likely mono-enriched, i.e., only one Pop. III progenitor contributed to enrich the birth clouds of these stars, as suggested by their extremely low metallicity and low [Mg/C] ($<-0.5$) ratios \citep{hartwig2018}. Thus, if the progenitors of the CEMP-no stars are CCSNe, faint SNe, and/or spinstars, their signatures are likely to be well-preserved in the Group II and III stars. However, the influence of the history of stochastic star formation, local inhomogeneous metal-mixing \citep[e.g.,][]{frebel2014,hartwig2019, simon2019}, and external enrichment/pollution by SN ejecta from their neighboring mini-halos  \citep[e.g.,][]{smith2015,chiaki2018} could be reflected in their stars as well. 

In contrast, dSph-like systems are much more massive than the UFDs, and contain substantially more baryonic gas, supporting extended star formation, resulting in strong chemical evolution in the system, as discussed in Section~\ref{sp}. This chemical evolution is well-reflected by the strong correlation among $A$(C), $A$(Na, Mg, Ba), and [Fe/H]. The majority of the next generation of stars might have formed internally (i.e., via internal pollution/self-enrichment) out of an already well-mixed interstellar medium enriched by various SNe from both Pop. III and Pop. II stars \citep[e.g.,][]{salvadori2015, debennassuti2017,sarmento2017,chiaki2018,sarmento2019}.

The most important inference from this comparison is that the mass of a host mini-halo plays a crucial role in influencing the CEMP group morphology of the stars formed there. All Group III CEMP-no stars may have formed in various low-mass mini-halos prior to accretion into the halo. We hypothesize that their overall level of $A$(C) reflects the role of limited dilution by their parent UFD-like systems.  The relatively high, plateau-like $A$(C) level of the Group III stars may constrain the lower mass limit of the accreted mini-halos, while the scatter in their observed $A$(C) could reflect differences in their local SFH, including influences arising from stochasticity, inhomogeneity, and external enrichment, as well as different mass progenitors. The strong correlation between $A$(C) and [Fe/H] among the Group II CEMP-no stars is clear evidence that they were likely formed in chemically evolved, more-massive parent systems prior to accretion into the halo. 
%\vfill\eject
\section{Conclusions}\label{conclusion}
We have investigated the origin of the CEMP-no group morphology of halo stars in the Yoon-Beers $A$(C)-[Fe/H] diagram of \citet{yoon2016}, in the context of a hierarchical Galactic assembly history, by comparing with the group morphology of CEMP-no stars presently found in dwarf satellites of the MW. Based on compiled literature data, we have confirmed that two distinct CEMP-no groups (Groups II and III) exist not only in the halo, but also in the satellite galaxies, supporting an accretion origin of the halo CEMP-no stars from their parent mini-halos.  While Group II CEMP-no stars are found in both the UFDs and the dSphs, the Group III CEMP-no stars appear to be found predominantly in the UFDs, indicating the importance of the parent mini-halo environment for CEMP-no group morphology. The anomalous Group I CEMP-no halo stars (with high $A$(C) but low [Ba/Fe]) may have had a similar origin as the Group III CEMP-no halo stars, although further study is required to validate this possibility. 

The relatively higher binary fraction ($47^{+15}_{
-14}$\%, \citealt{arentsen2019}) found among the Group I CEMP-no stars may indicate an association of their carbon enhancement with binary mass-transfer, however, their low absolute Ba abundances, which distinctively separate them from that of the majority of the CEMP-$s$ stars, is not accounted for by this scenario.  Future advances in extremely low-metallicity AGB models may also be able to address this phenomenon. We also speculate that highly C-enhanced birth clouds could have been enriched by  multiple sources of faint SNe and/or spinstars. Furthermore, the ubiquitously detected Ba (even a trace of it) in the atmospheres of the most metal-poor stars might have originated from either an $s$-process in spinstars and/or an $r$-process in neutron star mergers, collapsars, or magneto-rotationally driven jet supernovae.                               

The strong correlations between $A$(C,Na,Mg,Ba) and [Fe/H] among the Group II CEMP-no stars are signs that they were born in chemically evolved, massive dSph-like systems. In contrast, the higher levels of $A$(C) and the lack of dependence of $A$(C) on [Fe/H] and $A$(Na,Mg,Ba) for the Group III (and perhaps Group I) CEMP-no stars can be understood if low-mass, dark-matter dominated, chemically primitive systems are their birthplaces. In such environments, the influence of stochastic and truncated star-formation histories is maximal, arising from the effects of reionization/SNe feedback, local turbulence, and external enrichment by neighboring SNe as noted by many previous authors.  

Additional medium-, moderate-, and high-resolution spectroscopic analyses of Group II and III CEMP-no stars in both the MW halo and the satellite galaxies, and in particular for the presently small number of the anomalous Group I CEMP-no stars, are required in order to further our understanding of their origin, and in turn, the nature of the first stars. {\bf Furthermore, not only radial-velocity monitoring of the anomalous stars, but also additional evidence of mass transfer (e.g., spin-up of the receiving star), are required to confirm an association of their unusual abundances with a binary mass-transfer origin.} Using the Yoon-Beers diagram (Figure~\ref{ybd},~\ref{ybd2}) as ``a first-approximation'' diagnostic, we will be able to associate each CEMP sub-group (formation pathway) with a known CEMP sub-class (nucleosynthetic origin). High-resolution spectroscopic analyses for large samples of CEMP stars selected in a ``carbon-blind" (i.e., on the basis of [Fe/H] alone) fashion, now underway, are needed in order to firmly establish the frequencies of the various CEMP sub-classes as a function of [Fe/H]. These CEMP frequencies provide constraints on the initial mass function of Population III stars \citep{debennassuti2017} and, in turn, further our understanding of early Galactic chemical evolution history. 

\acknowledgments
{\bf We thank an anonymous referee for useful insights and a very constructive report, which led to a significant improvement in this work.} J.Y., T.C.B., and D.W.W. acknowledge partial support
from grant PHY 14-30152; Physics Frontier Center/JINA Center for the
Evolution of the Elements (JINA-CEE), awarded by the US National Science Foundation. D.T. participated this work as a Research Experiences for Undergraduates at the University of Notre Dame. This research made use of NASA's Astrophysics Data System, the
SIMBAD astronomical database, operated at CDS, Strasbourg, France, and
the SAGA database \citep{suda2008,yamada2013}
(http://sagadatabase.jp). 
\software{astropy \citep{astropy2013}, carbon evolutionary correction calculator (\citealt{placco2014c}, \url{https://vplacco.pythonanywhere.com/}), numpy \citep{numpy}, mplotlib \citep{hunter2007}    }

\bibliography{bibliography}

\end{document}